\documentclass[%
 reprint,
 amsmath,amssymb,
 aps,prx,
 twocolumn,
 showpacs,
 superscriptaddress,
 floatfix,
 showkeys,
 reprint,
 longbibliography,]
 {revtex4-2}

\usepackage{graphicx}% Include figure files
\usepackage{dcolumn}% Align table columns on decimal point
\usepackage{bm}% bold math
\usepackage{amsmath,amssymb}
\usepackage{hyperref}% add hypertext capabilities
\usepackage{natbib}
\usepackage{xcolor}
\usepackage[normalem]{ulem}
\usepackage{booktabs}
\usepackage{tabularx}
\usepackage[english]{babel}
\usepackage[utf8]{inputenc}
\usepackage{enumerate}
\usepackage{newunicodechar}
\usepackage{lipsum}

\usepackage[font=small,format=plain,labelfont=bf,
singlelinecheck=false,justification=raggedright,
labelsep=space,figurename=FIG.]{caption} 
\usepackage{ragged2e} % For justification
\begin{document}

\title{Platform for zero-field isolated skyrmions: 4$d$/Co atomic bilayers on Re(0001)}

\author{Moinak Ghosh}
\affiliation{Center for High-Performance Computing, Indian Institute of Science Education and Research Thiruvananthapuram, Thiruvananthapuram, Kerala 695551, India}

\author{Stefan Heinze}
%\email{heinze@physik.uni-kiel.de}
\affiliation{Institute of Theoretical Physics and Astrophysics, Christian-Albrechts-Universit$\ddot{a}$t zu Kiel, Leibnizstrasse 15, 24098 Kiel, Germany}
\affiliation{Kiel Nano, Surface, and Interface Science (KiNSIS), Christian-Albrechts-Universit$\ddot{a}$t zu Kiel, Christian-Albrechts-Platz 4, 24118 Kiel, Germany}

\author{Souvik Paul}
\email{souvikpaul@iisertvm.ac.in}
\affiliation{Center for High-Performance Computing, Indian Institute of Science Education and Research Thiruvananthapuram, Thiruvananthapuram, Kerala 695551, India}
\affiliation{School of Physics, Indian Institute of Science Education and Research Thiruvananthapuram, Thiruvananthapuram, Kerala 695551, India}

\date{\today}

\begin{abstract}
Using first-principles density functional theory (DFT) combined with atomistic spin simulations, we explore the possibility of realizing zero-field isolated skyrmions in three 4$d$/Co atomic bilayers -- Rh/Co, Pd/Co, and Ru/Co -- grown on the Re(0001) surface. Our investigation employs an extended atomistic spin model, which goes beyond the standard model by including the multi-spin higher-order exchange interactions (HOI) in addition to the Heisenberg pairwise exchange interaction, Dzyaloshinskii-Moriya interaction (DMI), and magnetocrystalline anisotropy energy (MAE). All magnetic interactions of the extended spin model are calculated using DFT. The phase diagram obtained from atomistic spin simulations based on this spin model for Rh/Co and Pd/Co on Re(0001) reveals that isolated skyrmions emerge spontaneously on the ferromagnetic background even in the absence of an external magnetic field. The radius of zero-field isolated skyrmions in Rh/Co/Re(0001) is around 6 nm, whereas the radius of those skyrmions in Pd/Co/Re(0001) is about 12 nm. Transition-state theory calculations show that the skyrmions are protected by substantial energy barriers, approximately 150 meV, which predominantly arise from DMI, with a small contribution from the HOI interactions. The height of the barriers suggests that skyrmions could be observed in low-temperature experiments. Based on this work, we propose 4$d$/Co bilayers on Re(0001) as a new platform to realize nanoscale zero-field isolated skyrmions.
\end{abstract}

%\keywords{DMFT, strongly correlated multiferroic, theoretical spectroscopy}
\maketitle

%\tableofcontents

\section{Introduction}
Topologically protected swirling spin structures in chiral magnets, such as magnetic skyrmions~\cite{nagaosa13,fert13,Bogdanov1989,Bogdanov1994,Roessler2006,binz2006,Muehlbauer2009,Yu2010a}, have raised high hopes to become the building block of next-generation spintronic devices~\cite{stuart2008,tomasello14,zhang2015,kang2016,tomasello2017,luo2018,nagaosa13}. This could be possible due to their nanoscale size, enhanced stability against external perturbations, and ultra-low current driven manipulation. With growing interest in magnetic skyrmions, researchers have proposed that these objects can also serve as information carriers in neuromorphic computing architectures~\cite{grollier2020,song2020} and as qubits in quantum technologies~\cite{psaroudaki2021,psaroudaki2022}.

The pioneering work of observing a nanoskyrmion lattice in an Fe monolayer on the Ir(111) surface using spin-polarized scanning tunneling microscopy (SP-STM) was the first realization of periodic arrangement of skyrmions at a transition-metal interface~\cite{Heinze2011}. Subsequently, isolated skyrmions were realized experimentally in various transition-metal ultrathin films~\cite{Romming2013,Romming2015,Hagemeister:15.1,kubetzka:17.1,Hanneken:15.1,meyer19} and in multilayer heterostructures~\cite{Boulle2016,Moreau-Luchaire2016,Woo2016,Soumya2017,chen2015}. The interfacial structure and composition of these materials can be easily altered, leading to tuning of the skyrmion properties. Moreover, these two-dimensional (2D) material systems can be integrated with existing electronic technologies, making this class of material suitable for potential applications~\cite{Dupe2016,Soumya2017}.

Skyrmions in transition-metal ultrathin films arise from the interplay among different magnetic interactions. These interactions include the Heisenberg pairwise exchange interaction, Dzyaloshinskii-Moriya interaction (DMI)~\cite{Dzyaloshinskii1957,Moriya1960}, and magnetocrystalline anisotropy energy (MAE). The DMI arises from spin-orbit coupling (SOC) in systems with broken inversion symmetry, e.g., at an interface, and is responsible for the chiral nature of skyrmions. Recently, it was demonstrated that the beyond Heisenberg pairwise exchange interaction, i.e., the higher-order multi-spin exchange interactions (HOI), such as the biquadratic, three-site four spin and four-site four spin interactions~\cite{takahashi77,macD88,hoffmann20}, play a crucial role for the stability of skyrmions in transition-metal films~\cite{paulhoi}.

In most transition-metal interfaces, isolated skyrmions are stabilized using the help of an external magnetic field. These systems show a shallow spin spiral energy minimum close to the ferromagnetic (FM) state. Isolated skyrmions appear at a magnetic field when the associated Zeeman energy compensates the spin spiral energy minimum. In rare cases, the strength of the magnetic interactions is such that isolated skyrmions spontaneously appear without the aid of an external magnetic field. Within the transition-metal ultrathin film family, only Rh/Co atomic bilayers on the surface of Ir(111) exhibit sub-10 nm isolated skyrmions at zero magnetic field, which were observed via SP-STM experiments and are stabilized due to exchange frustration~\cite{meyer19}. The field-free stabilization of skyrmions is particularly interesting from a device perspective, as it eliminates the requirement of a field-generating component and overcomes the challenge of applying magnetic field locally, thereby simplifying device integration and reducing energy consumption.

Recently, the Re(0001) surface was found as an experimentally promising substrate along with the previously explored Ir(111) and Rh(111) substrates. Various complex spin structures, such as isolated skyrmions, atomic-scale skyrmion lattices, and the noncollinear 3$Q$ state were reported in transition-metal films on the Re(0001) surface~\cite{spethmann2020,li2020,nickel2023,nickel2025}. An experimental study also reported the appearance of topological superconductivity in a magnet-superconductor hybrid system with the Re substrate~\cite{morales2019} and there are various theoretical proposals for such types of systems~\cite{bedow2020,mascot2021,nickel2025b}.

In an earlier theoretical study, we performed density functional theory (DFT) calculations for 4$d$/Co atomic bilayers on the surface of Re(0001) to explore their potential for the emergence of isolated skyrmions. Based on an effective spin model, which contains effective exchange interaction, effective DMI, and MAE, we proposed that four ultrathin films, i.e., fcc-Rh/Co, hcp-Pd/Co, fcc-Ru/Co, and fcc-Pd/Co on Re(0001), are promising candidates to host isolated skyrmions at zero magnetic field~\cite{paul2024}. We also suggested that, in another system, i.e., hcp-Rh/Co on Re(0001), isolated skyrmions could be stabilized only at a finite magnetic field. The two stackings of the 4$d$ overlayer on a magnetic monolayer in the DFT study are motivated by experimental observations in several transition-metal film systems~\cite{Romming2013,romming18,li2020}. However, the effective spin model neglected the HOI, and no atomistic spin simulations were carried out in that work to confirm any of the proposals. Among the four films, the first three films are studied in this paper, while the last system behaves differently, and will be addressed separately in a forthcoming publication~\cite{ncisk}. 

Here, we evaluate the magnetic interactions, such as the Heisenberg pairwise exchange, DMI, and MAE, as well as the HOI for fcc-Rh/Co, hcp-Pd/Co and fcc-Ru/Co on the Re(0001) surface using first-principles calculations based on DFT and construct an extended spin model with these interactions. Furthermore, using atomistic spin simulations based on the extended spin model, we explore the properties of isolated skyrmions, such as radius, collapse mechanism and stability, based on an atomistic spin Hamiltonian with the interaction constants evaluated from DFT.

Specifically, we calculate the energy dispersion of spin spirals without SOC for hcp-Pd/Co and fcc-Ru/Co on Re(0001), and with SOC for the three films using DFT. The energy dispersion of fcc-Rh/Co/Re(0001) without SOC was computed previously~\cite{paul2024}. The mapping of the DFT dispersion energies to the extended spin model yields the exchange and DMI interaction constants. We notice that the beyond nearest-neighbor exchange and DMI constants induce frustration in these films. The sign of the MAE constant indicates that the easy magnetization axis of all three films is out-of-plane. The three HOI interaction constants for fcc-Rh/Co and hcp-Pd/Co on Re(0001) range from relatively large to small values, with the biquadratic constant being positive, whereas the other two HOI constants, i.e., three-site four spin and four-site four spin, are negative. The strength of the HOI in fcc-Ru/Co/Re(0001) is negligible. The extended spin model used in this study does not overrule the conclusions obtained from the effective model, it merely modifies the skyrmion related properties. 

The magnetic phase diagram evaluated using atomistic spin simulations based on the extended spin model fully parameterized from DFT, discloses that isolated skyrmions appear spontaneously on the FM ground state at zero magnetic field for fcc-Rh/Co and hcp-Pd/Co on Re(0001). The radius of these zero-field skyrmions in fcc-Rh/Co/Re(0001) is around 6 nm, and it is approximately twice as large in hcp-Pd/Co/Re(0001). The energy barriers protecting isolated skyrmions in these two films are sufficiently high to allow their experimental observation at low temperatures. An analysis of the energy barriers reveals that the DMI is mostly responsible for the barriers along with a small contribution from the three-site four spin interaction. We could not stabilize isolated skyrmions in fcc-Ru/Co/Re(0001) at zero or finite magnetic fields due to the small DMI.

Based on our combined DFT and atomistic spin simulation approach, we propose 4$d$/Co bilayers on Re(0001) as a promising platform for realizing nanosized zero-field isolated skyrmions. We anticipate that our findings would stimulate experimental efforts to identify skyrmions in these ultrathin films. Moreover, these systems could serve as promising candidates for emergent topological superconductivity in magnet-superconductor hybrids, since Re becomes superconducting below 1.7~K.

This paper is organized in three sections, i.e., computational details (section II), results and discussions (section III), and conclusions (section IV). In the first section, we introduce the DFT method and the computational details used to calculate the electronic and magnetic properties, as well as atomistic spin simulations used to determine the skyrmion related properties. In the next section, we present and discuss our results. In the last section, we summarize our main findings and provide concluding remarks.

\section{\label{sec:compdet} Computational details}
\subsection{\label{sec:dftcalc} DFT calculations}
The electronic and magnetic structures of the ultrathin films were calculated using the $\textsc{fleur}$ code~\cite{fleur} that is based on the full-potential linearized augmented plane wave (FP-LAPW) method. This code is ideal for calculating the collinear as well as noncollinear magnetic properties including SOC for 2D materials, such as transition-metal ultrathin films~\cite{kurz2004,heide2009,Zimmermann2014}. 

We have chosen the experimental lattice constant of bulk Re, i.e., $a= 2.761$ \AA, as the in-plane lattice constant, and bulk layer separation $c/2= 2.228$ \AA~as the interlayer distances of the ultrathin films~\cite{re}. To mimic the experimental geometry, we considered 4$d$/Co atomic bilayers on top of nine layers of the Re(0001) surface, which models the substrate. The Co layer was chosen with hcp stacking on the Re(0001) surface, while different stacking orders (fcc or hcp) were chosen for the 4$d$ transition-metal overlayers. The geometrical optimization was performed to determine the top three interlayer distances which were reported in Ref.~\cite{paul2024} and used here. In this work, we used the local density approximation for the exchange-correlation part of the potential as parameterized by Vosko, Wilk, and Nusair~\cite{vwn}. The energy cutoff of the plane wave was chosen as $k_{\rm max}=4.0$~a.u.$^{-1}$ and a dense $k$-mesh of 44$\times$44 points was used in the full 2D Brillouin zone (BZ) to achieve good convergence.

The standard way to search for a noncollinear ground state in an ultrathin film is to calculate the energy dispersion of spin spirals in the 2DBZ. Following this approach, we calculated the total DFT energy of homogeneous flat spin spirals as a function of the propagation vector $\mathbf{q}$ along the two high-symmetry directions of the 2DBZ. We used the generalized Bloch theorem, which allows us to perform spin spiral calculations within the chemical unit cell, as implemented in the $\textsc{fleur}$ code~\cite{kurz2004}.

Thereafter, the total energies of homogeneous spin spirals were mapped onto the Heisenberg Hamiltonian to extract the pairwise exchange interaction constants. We calculated the exchange constants up to the 10-th nearest-neighbor to understand the stability of the noncollinear state compared to the FM state. The Heisenberg Hamiltonian can be expressed as
\begin{align} \label{eq1}
\mathcal{H}_{\rm ex} =- \sum_{ij} J_{ij} \ (\textbf{m}_{i}\cdot\textbf{m}_{j})
\end{align}
where the pairwise exchange interaction constants are denoted as $J_{ij}$ and $\textbf{m}_{i}$ ($\textbf{m}_{j}$) is the unit vector along the magnetization direction at the site $i$ ($j$). The positive (negative) value of $J_{ij}$ indicates that an FM (antiferromagnetic) configuration is favored. 

The right (clockwise) and left (anticlockwise) rotating spin spirals are degenerate within the Heisenberg model. However, it is lifted when SOC is included. SOC brings in two contributions, namely DMI and MAE. The first one arises due to broken inversion symmetry at the interface. Since the DMI energy is relatively small compared to the total energy, we computed it within the first-order perturbation theory based on the self-consistent spin spiral state~\cite{heide2009,Zimmermann2014}. The DMI Hamiltonian can be expressed as~\cite{Dzyaloshinskii1957,Moriya1960} 
\begin{align} \label{eq2}
\mathcal{H}_{\rm DMI} =- \sum_{ij} \boldsymbol{D}_{ij} \cdot(\textbf{m}_{i}\times\textbf{m}_{j})
\end{align}
where $\boldsymbol{D}_{ij}$ is the DMI vector. The symmetry of the films permits an out-of-plane component of the DMI vector~\cite{vida2016}, however this component is typically small in such films. Therefore, we have chosen the DMI vectors in the plane of the film and perpendicular to the line joining $\textbf{m}_{i}$ and $\textbf{m}_{j}$, which promotes cycloidal spin spirals. The DMI lowers the energy of spin spirals with preferred rotational sense, whether clockwise or anticlockwise. With our sign convention, the positive (negative) value of the DMI constant indicates that clockwise (anticlockwise) homogeneous cycloidal spin spirals are favored.

The second term arising from SOC is the MAE, which was evaluated self-consistently based on the energy difference between the in-plane and out-of-plane magnetization directions within the second variation approach~\cite{Li1990}. We define the MAE constant as $K_{\mathrm{MAE}}$= $E_{\parallel}$ $-$ $E_{\perp}$, where $E_{\parallel}$ and $E_{\perp}$ denote the total energies for an in-plane and out-of-plane magnetization direction, respectively. Thus, the positive (negative) value of the MAE constant indicates an out-of-plane (in-plane) easy magnetization axis.

The HOI terms arise from the fourth-order perturbative expansion of the hopping parameter over the Coulomb term in the Hubbard model~\cite{hoffmann20}. This expansion results in a two-site four spin interaction, widely known as the biquadratic interaction, a three-site four spin interaction~\cite{macD88,hubbard3}, and a four-site four spin interaction~ \cite{hoffmann20}. Due to the fourth-order perturbation, we restrict the spin model to nearest-neighbor interaction terms for the three HOI. The Hamiltonians of the three HOI can be expressed as
\begin{gather}
\mathcal{H}_{\mathrm{biquad}} = - B_1 \sum_{\langle ij \rangle} (\textbf{m}_{i} \cdot \textbf{m}_{j})^2 \\
\mathcal{H}_{\mathrm{3-site}} = - 2 Y_1 \sum_{\langle ijk \rangle} (\textbf{m}_{i} \cdot \textbf{m}_{j}) (\textbf{m}_{j} \cdot \textbf{m}_{k}) \\
\mathcal{H}_{\mathrm{4-site}} = - K_1 \sum_{\langle ijkl \rangle} [(\textbf{m}_{i} \cdot \textbf{m}_{j}) (\textbf{m}_{k} \cdot \textbf{m}_{l}) \nonumber \\
	+(\textbf{m}_{i} \cdot \textbf{m}_{l}) (\textbf{m}_{j} \cdot \textbf{m}_{k})-(\textbf{m}_{i} \cdot \textbf{m}_{k}) (\textbf{m}_{j} \cdot \textbf{m}_{l})]
\end{gather}
where $B_1$, $Y_1$, and $K_1$ are the nearest-neighbor biquadratic, three-site four spin and four-site four spin constants, respectively, and the nearest-neighbor approximation is indicated by $\langle...\rangle$. For the 2D hexagonal lattice considered here, the summation $\langle ijkl \rangle$ runs over the four sites of a diamond, whereas the sum $\langle ijk \rangle$ runs over the three vertices of a triangle~\cite{paulhoi}. The last two interaction terms are abbreviated in the Hamiltonian as 3-site and 4-site, respectively. 

To compute these three HOI constants, we considered three special multi-$Q$ states: two 2D collinear up-up-down-down ($uudd$) or double row-wise antiferromagnetic (AFM) states along the $\overline{\Gamma \mathrm{K}}$ and $\overline{\Gamma \mathrm{M}}$ directions of the 2DBZ~\cite{hardrat2009,Kronelein2018,romming18}, and one 3D noncollinear state at the $\overline{\mathrm{M}}$ point, known as the 3$Q$ state~\cite{pkurz,spethmann2020,Nickel2023b}. These special magnetic states are obtained from the superposition of spin spirals corresponding to symmetry equivalent $\textbf{q}$ points in the 2DBZ. The HOI were evaluated from the total energy difference between the multi-$Q$ and the corresponding single-$Q$ states without SOC as follows
\begin{gather} 
	B_{1}=\frac{3}{32} \Delta E_{\overline{\mathrm{M}}}^{3Q} - \frac{1}{8} \Delta E_{\overline{\mathrm{M}}/2}^{uudd} \label{eq6} \\
	Y_{1}=\frac{1}{8}(\Delta E_{3\overline{\mathrm{K}}/4}^{uudd} - \Delta E_{\overline{\mathrm{M}}/2}^{uudd}) \label{eq7} \\
	K_{1}=\frac{3}{64} \Delta E_{\overline{\mathrm{M}}}^{3Q} + \frac{1}{16} \Delta E_{3{\overline{\mathrm{K}}}/4}^{uudd} \label{eq8}
\end{gather}
where the energy differences are given by
\begin{gather}
	\Delta E_{\overline{\mathrm{M}}}^{3Q}= E_{\overline{\mathrm{M}}}^{3Q} - E_{\overline{\mathrm{M}}}^{SS}	 \label{eq9} \\
	\Delta E_{\overline{\mathrm{M}}/2}^{uudd}= E_{\overline{\mathrm{M}}/2}^{uudd} - E_{\overline{\mathrm{M}}/2}^{SS}  \label{eq10} \\
	\Delta E_{3\overline{\mathrm{K}}/4}^{uudd}= E_{3{\overline{\mathrm{K}}}/4}^{uudd} - E_{3{\overline{\mathrm{K}}}/4}^{SS} \label{eq11} \\
	\nonumber
\end{gather}
where $\Delta E_{\overline{\mathrm{M}}}^{3Q}$ is the energy difference between the 3$Q$ state and the spin spiral state at the $\overline{\mathrm{M}}$ point, $\Delta E_{\overline{\mathrm{M}}/2}^{uudd}$ is the energy difference between the $uudd$ state and the spin spiral state at $\overline{\mathrm{M}}/2$ along the $\overline{\Gamma \mathrm{M}}$ direction, and $\Delta E_{3\overline{\mathrm{K}}/4}^{uudd}$ is the energy difference between the $uudd$ state and the spin spiral state at $3\overline{\mathrm{K}}/4$ along the $\overline{\Gamma \mathrm{K}}$ direction.

\subsection{\label{sec:asds} Atomistic spin simulations}
We studied the relaxation of spin spirals, skyrmion lattices, and isolated skyrmions and computed their energies in the presence of external magnetic fields through atomistic spin simulations based on the Landau-Lifshitz equation expressed as
\begin{gather} \label{eq:llg}
\hbar \frac{d\textbf{m}_{i}}{dt}=\frac{\partial \mathcal{H}}{\partial{\textbf{m}}_{i}}\times\textbf{m}_{i}-\alpha\left(\frac{\partial \mathcal{H}}{\partial{\textbf{m}}_{i}}\times\textbf{m}_{i}\right)\times\textbf{m}_{i}
\end{gather}
where $\hbar$ is the reduced Planck constant, $\alpha$ is the damping parameter, and the Hamiltonian $\mathcal{H}$ is given by
\begin{gather}
\mathcal{H}= \mathcal{H}_{\rm ex} + \mathcal{H}_{\rm DMI} + \mathcal{H}_{\rm MAE} +  \mathcal{H}_{\rm Zeeman} + \nonumber \\ \mathcal{H}_{\mathrm{biquad}} + \mathcal{H}_{\mathrm{3-site}} + \mathcal{H}_{\mathrm{4-site}}
\end{gather}
Here $\mathcal{H}_{\rm ex}$, $\mathcal{H}_{\rm DMI}$, $\mathcal{H}_{\mathrm{biquad}}$, $\mathcal{H}_{\mathrm{3-site}}$, $\mathcal{H}_{\mathrm{4-site}}$ are given in Eqs.~(1-5), respectively, and $\mathcal{H}_{\rm MAE}$, $\mathcal{H}_{\rm Zeeman}$ can be written as
\begin{gather}
\mathcal{H}_{\rm MAE}=	- \sum_{i} K_{\mathrm{MAE}}(m^{z}_{i})^2 \\
\mathcal{H}_{\rm Zeeman}= - \sum_{i} \mu_{s} \textbf{B} \cdot \textbf{m}_{i}
\end{gather}
Here, $K_{\mathrm{MAE}}$ is the MAE constant, where a positive (negative) sign indicates an out-of-plane (in-plane) easy magnetization axis. $B$ is the external magnetic field and $\mu_{s}$ is the total magnetic moment per unit cell of the film obtained via DFT. Note that all the interaction constants in the Hamiltonian of Eq.~(13) were completely determined from DFT. We do not actually study spin dynamics via the Landau-Lifshitz equation, it is only used here for relaxing the spin structures to their minimum energy configuration.

The Landau-Lifshitz equation was solved using a semi-implicit integration scheme proposed by Mentink \textit{et al}.~\cite{mentink2000} in a 2D hexagonal lattice of 150$\times$150 spins. We used a damping constant between the range 0.25 to 0.50, a time step of 0.1 fs, and around 2-3 million iterations to ensure good relaxation.

The collapse mechanism of the isolated skyrmions and therefore their energy barriers were calculated using the GNEB method~\cite{bessarab2015,malottki2017a}. The method creates a discrete starting path joining the initial (isolated skyrmions) and final (FM) states through intermediate spin configurations, called images, which are separated by a spring force. The goal of the method is to bring the starting path to the minimum energy path (MEP) via relaxing the intermediate images systematically. This relaxation process is performed according to a force projection scheme. Within this scheme, the force is calculated at each image, and its perpendicular component is retained, while the tangential component is replaced by the spring force, which keeps the images at a uniform distance. The maximum energy point along the minimum energy path is called the saddle point, whose energy with reference to the initial and final states is defined as the annihilation and creation barriers, respectively.

\begin{figure*}[!htbp]
	\includegraphics[scale=1.0]{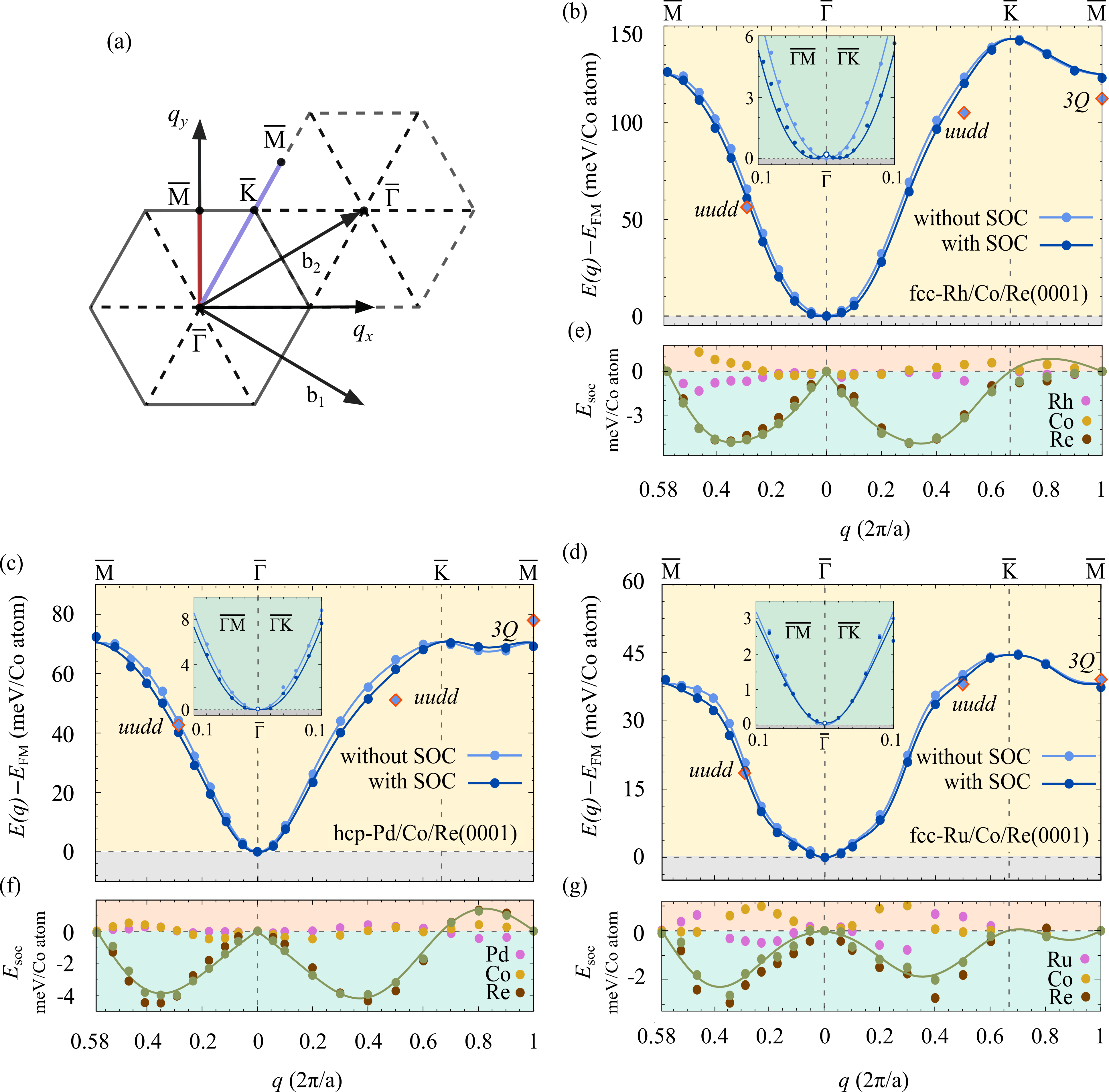}
	\centering
	\caption{\justifying (a) Two-dimensional Brillouin zone (2DBZ) displaying two high-symmetry directions $\overline{\Gamma \mathrm{KM}}$ (blue) and $\overline{\Gamma \mathrm{M}}$ (red). The reciprocal vectors ($\mathbf{b}_1$ and $\mathbf{b}_2$) are also shown. Energy dispersion $E$(\textbf{q}) of homogeneous flat spin spirals along the $\overline{\Gamma \mathrm{KM}}$ and $\overline{\Gamma \mathrm{M}}$ directions of the 2DBZ without (light blue) and with (dark blue) SOC for (b) fcc-Rh/Co/Re(0001), (c) hcp-Pd/Co/Re(0001), and (d) fcc-Ru/Co/Re(0001). Filled circles present DFT data and solid lines are fits to the pairwise Heisenberg exchange (Eq.~(1)) (light blue) and pairwise Heisenberg exchange plus the DMI (Eq.~(2)) (dark blue) spin models. Filled diamonds, highlighted by solid red border, present the DFT total energies of multi-$Q$ states ($uudd$ and 3$Q$), evaluated without SOC, and are shown at the $\textbf{q}$ points of the corresponding single-$Q$ (spin spiral) states. Insets show $E$(\textbf{q}) in the vicinity of the $\overline{\Gamma}$ point ($\mid$$\textbf{q}$$\mid \leq 0.1\times\frac{2\pi}{a}$) without and with SOC. Note that the dispersions with SOC contain the effect of DMI and MAE. The MAE shifts the energy dispersion ($E(\textbf{q})$) by $K_{\textrm{MAE}}$/2 along the positive energy axis with respect to the FM state ($\overline{\Gamma}$ point). Total energy contribution due to SOC (green) and individual contributions from 4$d$ overlayers (pink), i.e., (e) fcc-Rh, (f) hcp-Pd, and (g) fcc-Ru, Co (yellow) and Re substrate (brown) along the two high-symmetry directions $\overline{\Gamma \mathrm{KM}}$ and $\overline{\Gamma \mathrm{M}}$ of the 2DBZ. Solid circles denote DFT data and solid green lines are fits to the DMI Hamiltonian (Eq.~(2)). The full energy dispersion of fcc-Rh/Co/Re(0001) without SOC (light blue) in (b) is taken from Ref.~\cite{paul2024}}. 
	\label{fig:f1}
\end{figure*}

\section{\label{sec:resdiss} Results and discussions}
\subsection{DFT calculations}
\subsubsection{\label{sec:engdis} Energy dispersion and multi-$Q$ states}

Based on DFT calculations, we have previously found that the energy dispersion, $E$(\textbf{q}), of spin spirals, including SOC, for the atomic bilayers of fcc-Rh/Co, hcp-Pd/Co, and fcc-Ru/Co on Re(0001) is quite flat in the vicinity of the $\overline{\Gamma}$ point ($\mid$$\textbf{q}$$\mid \leq 0.1\times\frac{2\pi}{a}$) along the two high-symmetry directions $\overline{\Gamma \mathrm{KM}}$ and $\overline{\Gamma \mathrm{M}}$ of the 2DBZ (Fig.~\ref{fig:f1}(a)) \cite{paul2024}. A similar slow rise of the energy dispersions was observed in Rh/Co atomic bilayers on the Ir(111) surface, where isolated skyrmions are stabilized even in the absence of a magnetic field, and observed via SP-STM measurements~\cite{meyer19}. These facts suggest that the three ultrathin films mentioned above are promising candidates to host isolated skyrmions. Furthermore, the full energy dispersion of fcc-Rh/Co/Re(0001) without SOC along the $\overline{\Gamma \mathrm{KM}}$ and $\overline{\Gamma \mathrm{M}}$ directions was reported in Ref.~\cite{paul2024}.

Here, we calculate the full energy dispersion via DFT along the $\overline{\Gamma \mathrm{KM}}$ and $\overline{\Gamma \mathrm{M}}$ directions, both without and with SOC for hcp-Pd/Co and fcc-Ru/Co on Re(0001), and with SOC for fcc-Rh/Co/Re(0001) (Figs.~\ref{fig:f1}(b-d)). For comparison, the full energy dispersion without SOC for fcc-Rh/Co/Re(0001), taken from Ref.~\cite{paul2024}, is also shown.

The energy dispersion of fcc-Rh/Co/Re(0001), neglecting SOC, follows a parabolic ($q^2$) dependence close to the $\overline{\Gamma}$ point, reflecting the dominance of the nearest-neighbor exchange interaction (Fig.~\ref{fig:f1}(b)). As the wave vector $q$ increases, the interactions beyond the nearest-neighbor influence the energy dispersion, leading to a deviation from the parabolic shape. 

The energy dispersion of fcc-Rh/Co/Re(0001) rises sharply with $q$ along the $\overline{\Gamma \mathrm{KM}}$ direction and reaches a maximum value at the $\overline{\mathrm{K}}$ point, which corresponds to the N\'eel state. Thereafter, the energy gradually decreases to the $\overline{\mathrm{M}}$ point, which indicates the row-wise antiferromagnetic (RW-AFM) state. The energy of spin spirals also rises quickly along the $\overline{\Gamma \mathrm{M}}$ direction and reaches a maximum value at the edge of the 2DBZ, i.e., at the $\overline{\mathrm{M}}$ point. Clearly, the energy of the FM state ($\overline{\Gamma}$ point) is lowest among all.

In the inset of Fig.~\ref{fig:f1}(b), we focus on the energy dispersion close to the $\overline{\Gamma}$ point ($\mid$$\textbf{q}$$\mid \leq 0.1\times\frac{2\pi}{a}$). The effect of SOC is clearly visible. The dispersion becomes flatter along both the high-symmetry directions $\overline{\Gamma \mathrm{KM}}$ and $\overline{\Gamma \mathrm{M}}$ of the 2DBZ as SOC is included. However, the FM state has the lowest energy even after adding SOC, and is therefore considered as the ground state of the film.

The energy dispersion of hcp-Pd/Co and fcc-Ru/Co on Re(0001) (Figs.~\ref{fig:f1}(c-d)), neglecting SOC, behaves qualitatively in a similar manner to that of fcc-Rh/Co/Re(0001). However, the latter film exhibits the smallest curvature (inset of Fig.~\ref{fig:f1}(d)), while the former film shows the highest curvature (inset of Fig.~\ref{fig:f1}(c)) among the three films considered. As a result, the dispersion of the spin spiral around the $\overline{\Gamma}$ point of fcc-Ru/Co/Re(0001) lies lowest in energy and that of hcp-Pd/Co/Re(0001) lies highest in energy, while that of fcc-Rh/Co/Re(0001) lies in between the two. However, at large $q$, the ordering in the energy dispersion between hcp-Pd/Co and fcc-Rh/Co on Re(0001) is reversed. The dispersion curve of fcc-Rh/Co/Re(0001) rises more sharply than hcp-Pd/Co/Re(0001) and reaches a maximum energy of nearly 145 meV, while the maximum energy attained by the dispersion curve of hcp-Pd/Co/Re(0001) is approximately 70 meV. The energy of the spin spiral dispersion of fcc-Ru/Co/Re(0001) always remains the lowest among the three. The insets of Figs.~\ref{fig:f1}(c-d) shows a similar trend as of the fcc-Rh/Co/Re(0001), and the FM state is also the ground state of the other two films.

To evaluate the HOI constants, we calculate the total energy of three special multi-$Q$ states -- the two $uudd$ states~\cite{hardrat2009,Kronelein2018,romming18} and a 3$Q$ state~\cite{pkurz,spethmann2020,Nickel2023b} -- for each film via DFT, which are shown at the positions of corresponding single-$Q$ states in Figs.~\ref{fig:f1}(b-d). The energy of the three states for fcc-Rh/Co/Re(0001) are lower than the respective single-$Q$ states. For hcp-Pd/Co/Re(0001), the 3$Q$ state lies higher in energy, while the $uudd$ state along the $\overline{\Gamma \mathrm{KM}}$ direction lies lower in energy compared to the single-$Q$ state. The $uudd$ state along the $\overline{\Gamma \mathrm{M}}$ direction is located almost on top of the single-$Q$ state. Overall, the energy difference between the three multi-$Q$ and single-$Q$ states for fcc-Ru/Co/Re(0001) is smaller than that of the other two films. However, the 3$Q$ state lies slightly higher in energy compared to the single-$Q$ state, while the two $uudd$ states remain lower in energy than their respective single-$Q$ states by a small amount.

\begin{figure}[!htbp]
	\includegraphics[scale=1.0]{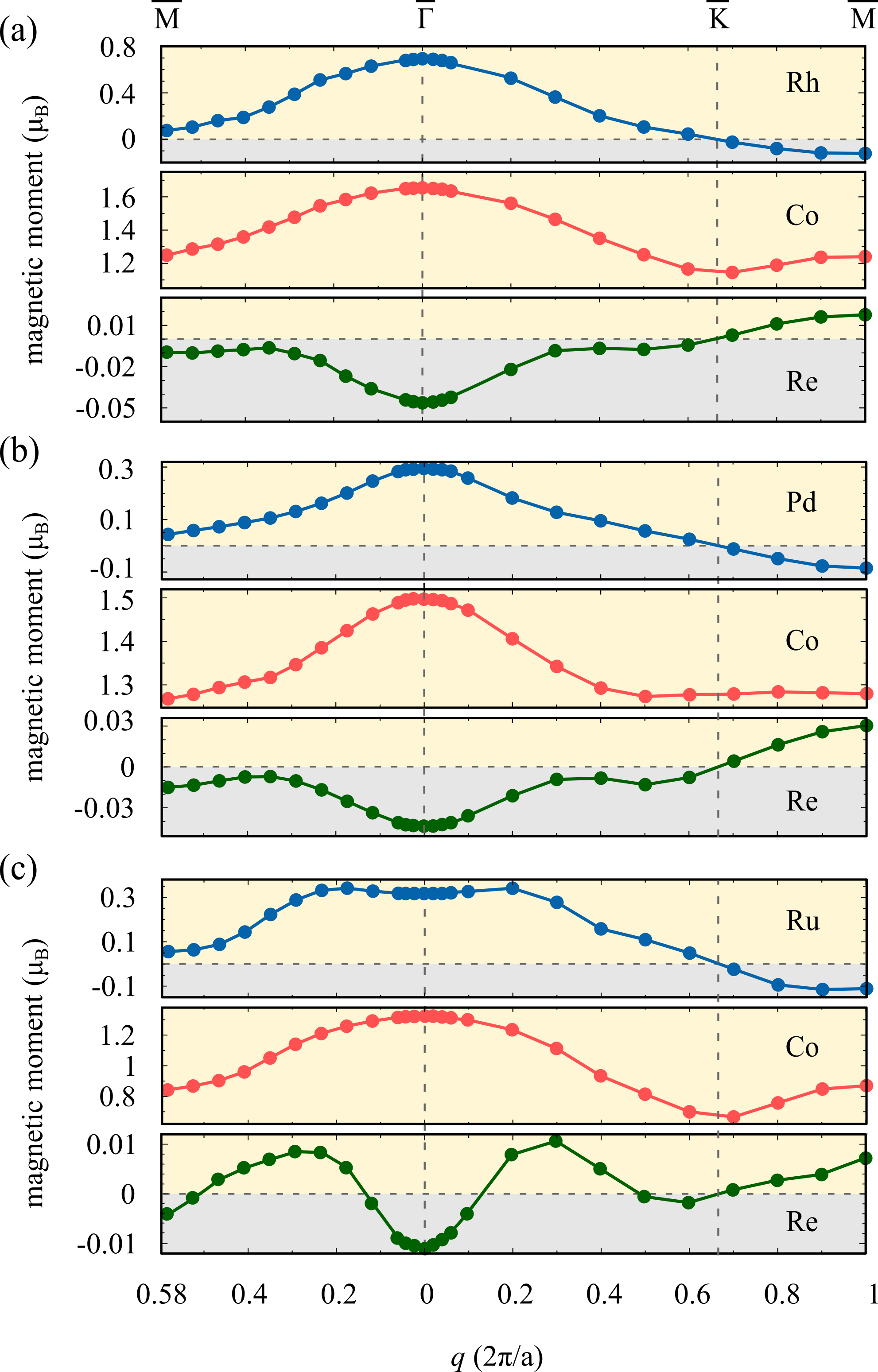}
	\centering
	\caption{\justifying Variation of the magnetic moments in (a) fcc-Rh/Co/Re(0001), (b) hcp-Pd/Co/Re(0001), and (c) fcc-Ru/Co/Re(0001) for spin spirals along the two high-symmetry directions $\overline{\Gamma \mathrm{KM}}$ and $\overline{\Gamma \mathrm{M}}$ of the 2DBZ shown in Figs.~\ref{fig:f1}(b-d), respectively. Top, middle, and bottom panels of (a) show the magnetic moment of the Rh, Co, and the first Re layers, respectively; those of (b) show the Pd, Co, and the first Re layers, respectively, and those of (c) show of the Ru, Co, and the first Re layers, respectively.} 
	\label{fig:f2}
\end{figure}

Next, we discuss the effects due to SOC, i.e., the MAE and the DMI. The MAE was calculated for the FM state for all three systems in Ref.~\cite{paul2024}. The easy magnetization axis of all three films is perpendicular to the film plane, i.e., they exhibit an out-of-plane easy axis. The value of the MAE is 0.60 meV per Co atom for fcc-Rh/Co/Re(0001), and it is nearly 5 times smaller for the other two systems (Table~\ref{tab:table1}).

The DMI arises due to SOC in combination with the inversion symmetry breaking at the surface. Note that, in accordance with the symmetry of the film, the DMI promotes cycloidal spin spirals. This term is significant for the heavy 4$d$ and 5$d$ transition metals, such as Rh, Pd, Ru, and Re, which possess a large SOC due to their high nuclear charge. The contribution of DMI to the energy dispersion of cycloidal spin spirals for the three films is shown in Figs.~\ref{fig:f1}(e-g). 

For fcc-Rh/Co/Re(0001), the DMI lowers the energy of spin spirals by a small amount at low and moderate $q$ values along the $\overline{\Gamma \mathrm{KM}}$ direction and beyond the $\overline{\mathrm{K}}$ point, the contribution becomes negligible. The DMI contribution also lowers the spin spiral energy by a small amount along the $\overline{\Gamma \mathrm{M}}$ direction. However, the ground state of the system is still the FM state (inset of Fig.~\ref{fig:f1}(b)). The energy of the spin spiral including SOC for hcp-Pd/Co and fcc-Ru/Co on Re(0001) follows the same trend as that observed for fcc-Rh/Co/Re(0001). Both films exhibit an FM ground state as well (insets of Figs.~\ref{fig:f1}(c,d)).

\begin{table*}[!thbp]
	\centering
	\caption{\justifying Effective exchange interaction constant ($J_{\mathrm{eff}}$), $i$-th nearest-neighbor pairwise exchange constant ($J_{i}$), three HOI constants, i.e., biquadratic ($B_{1}$), three-site four spin ($Y_{1}$), and four-site four spin ($K_{1}$), effective DMI constant ($D_{\mathrm{eff}}$), $i$-th nearest-neighbor DMI ($D_{i}$) constants, MAE constant ($K_{\mathrm{MAE}}$), energy difference between the multi-$Q$ and single-$Q$ states ($\Delta E_{3\overline{\mathrm{K}}/4}^{uudd}$, $\Delta E_{\overline{\mathrm{M}}/2}^{uudd}$, and $\Delta E_{\overline{\mathrm{M}}}^{3Q}$), and total magnetic moment ($\mu_{\mathrm{s}}$) are obtained from DFT for fcc-Rh/Co, hcp-Pd/Co, and fcc-Ru/Co on Re(0001), and used in the atomistic spin simulations. The positive value of $K_{\mathrm{MAE}}$ indicates an out-of-plane easy magnetization axis. The effective exchange constant, effective DMI constant, and MAE for all three films are taken from Ref.~\cite{paul2024}. The pairwise exchange constants for fcc-Rh/Co/Re(0001) are also taken from Ref.~\cite{paul2024}. The magnetic interaction constants and the energy difference ($\Delta E$) are given in meV, while the magnetic moment is given in Bohr magneton ($\mu_{\mathrm{B}}$).} 
	\label{tab:table1}
	\begin{ruledtabular}
		\begin{tabular}{cccccccccccccccc}
			System &$J_{\mathrm{eff}}$ & $J_{1}$ & $J_{2}$ & $J_{3}$ & $J_{4}$ & $J_{5}$ & $J_{6}$ & $J_{7}$ & $J_{8}$ & $J_{9}$ & $J_{10}$ & $B_{1}$ & $Y_{1}$ &$K_{1}$ &$\mu_{\mathrm{s}}$\\
			\colrule
			fcc-Rh &12.67 &16.85 &1.56 &0.18 &0.03 &$-$0.42 &{---} &{---} &{---} &{---} &{---} &0.15 &$-$1.13 &$-$1.65 &2.3 \\
			hcp-Pd &15.12 &8.46 &2.77 &$-$0.37 &0.29 &$-$0.18 &$-$0.01 &$-$0.09 &{---} &{---} &{---} &1.01 &$-$1.50 &$-$0.44 &1.8 \\
			fcc-Ru &6.90 &5.01 &0.32 &0.10 &$-$0.29 &$-$0.32 &$-$0.11 &$-$0.02 &0.07 &0.08 &0.12 &0.45 &0.00 &$-$0.06 &1.6 \\
			\colrule
			\hline
			System &$\Delta E_{3\overline{\mathrm{K}}/4}^{uudd}$ &$\Delta E_{\overline{\mathrm{M}}/2}^{uudd}$ &$\Delta E_{\overline{\mathrm{M}}}^{3Q}$ &$D_{\mathrm{eff}}$ & $D_{1}$ & $D_{2}$ & $D_{3}$ & $D_{4}$ & $D_{5}$ & $K_{\mathrm{MAE}}$ & ${}$ & ${}$ & ${}$ & ${}$ & ${}$\\
			\colrule
			fcc-Rh &$-$18.32 &$-$9.28 &$-$10.77 &1.17 &1.47 &$-$0.08 &$-$0.12 &0.03 &0.04 &0.60 & ${}$ & ${}$ & ${}$ & ${}$ & ${}$ \\
			hcp-Pd &$-$13.54 &$-$1.51 &8.73 &0.70 &1.19 &$-$0.22 &$-$0.07 &{---} &{---} &0.14 & ${}$ & ${}$ & ${}$ & ${}$ & ${}$ \\
			fcc-Ru &$-$2.26 &$-$2.24 &1.82 &0.07 &0.55 &0.00 &$-$0.23 &0.02 &$-$0.04 &0.11 & ${}$ & ${}$ & ${}$ & ${}$ & ${}$ \\
		\end{tabular}
	\end{ruledtabular}
\end{table*}

To analyze the DMI energy in more detail, we decompose it into the contributions from different layers as displayed in Figs.~\ref{fig:f1}(e-g), for fcc-Rh/Co, hcp-Pd/Co, and fcc-Ru/Co on Re(0001), respectively. The total DMI contribution to the energy dispersion of fcc-Rh/Co/Re(0001) is noticeable along the $\overline{\Gamma \mathrm{K}}$ portion of the $\overline{\Gamma \mathrm{KM}}$ direction and along the $\overline{\Gamma \mathrm{M}}$ direction (Fig.~\ref{fig:f1}(b)). The negative DMI energy in these two regions indicates that the clockwise cycloidal spin spirals are preferred, which lowers the energy of the spin spirals calculated without SOC. The minimum of the total DMI energy occurs at around $q =0.35\times\frac{2\pi}{a}$ along the $\overline{\Gamma \mathrm{KM}}$ direction with a depth of approximately 5 meV per Co atom relative to the $q=0$ state (Fig.~\ref{fig:f1}(e)). At small $q$ values, the contributions of the Rh and Co layers are negligible. However, at large $q$ values, the negative contribution of the Rh layer almost cancels the positive contribution of the Co layer. As a result, the total DMI energy follows the contribution of the Re substrate.

The total DMI energy of hcp-Pd/Co/Re(0001) behaves in a similar way to that of fcc-Rh/Co/Re(0001) (Fig.~\ref{fig:f1}(f)). The layerwise decomposition shows that the DMI energy of the Pd and Co layers is negligible, and the total DMI energy is governed by the contribution of the Re substrate. The minimum occurs at around $q=0.38\times\frac{2\pi}{a}$ along the $\overline{\Gamma \mathrm{KM}}$ direction, which is nearly 4.2 meV per Co atom lower than the FM state ($q=0$). 

The total DMI energy of fcc-Ru/Co/Re(0001) follows the same trend as those of the other two films. The minimum occurs at around $q=0.38\times\frac{2\pi}{a}$ along the $\overline{\Gamma \mathrm{M}}$ direction with a depth of nearly 2.2  meV per Co atom relative to the FM state. The DMI energy of fcc-Ru/Co/Re(0001), arising from the Ru and Co layers, behaves in a different way (Fig.~\ref{fig:f1}(g)) than the other two films. At small and intermediate $q$, the Ru and Co layers exhibit noncancelable positive and negative contributions, respectively. As $q$ increases further, the DMI energy of the Co layer becomes zero and the Ru layer first becomes positive and then vanishes. As a result, the total DMI energy does not follow the Re contributions, and its value is smaller than those of the other two films.

\subsubsection{Magnetic moments and interaction constants}
The total magnetic moment of fcc-Rh/Co/Re(0001) in the FM state is 2.3 $\mu_{\mathrm{B}}$, which is nearly 1.3 times lower for the other two films (Table~\ref{tab:table1}). The magnetic moments of the top two layers and the first Re layer for flat spin spirals, shown in Fig.~\ref{fig:f1}(b-d), are displayed in Fig.~\ref{fig:f2} along the two high-symmetry directions $\overline{\Gamma \mathrm{KM}}$ and $\overline{\Gamma \mathrm{M}}$ of the 2DBZ for the three films. Overall, the magnetic moment of the Co monolayer in the three films varies comparatively more than that of the Fe monolayer in other TM ultrathin films along the two high-symmetry directions of 2DBZ. The variation in the magnetic moment of the Co layer is nearly 0.5 $\mu_{\mathrm{B}}$ for fcc-Rh/Co/Re(0001) (Fig.~\ref{fig:f2}(a)), nearly 0.3 $\mu_{\mathrm{B}}$ for hcp-Pd/Co/Re(0001) (Fig.~\ref{fig:f2}(b)), and nearly 0.6 $\mu_{\mathrm{B}}$ for fcc-Ru/Co/Re(0001) (Fig.~\ref{fig:f2}(c)). The induced magnetic moments of the adjacent $4d$ layers, i.e., fcc-Rh, hcp-Pd, and fcc-Ru, also vary in response to the magnetic layer by about 0.8 $\mu_{\mathrm{B}}$ for fcc-Rh/Co/Re(0001) (Fig.~\ref{fig:f2}(a)), 0.4 $\mu_{\mathrm{B}}$ for hcp-Pd/Co/Re(0001) (Fig.~\ref{fig:f2}(b)), and fcc-Ru/Co/Re(0001) (Fig.~\ref{fig:f2}(c)). Such a large variation in magnetic moments of the Rh and Co layers was reported in Rh/Co/Ir(111) ultrathin films~\cite{meyer19,meyer2020}. The substantial variation of magnetic moment observed in these films cannot be described by the Heisenberg model (Eq.~(1)), which assumes a fixed magnetic moment. 

However, the magnetic moments of the Rh and Co layers in fcc-Rh/Co/Ir(111) close to the FM state ($\overline{\Gamma}$ point) remain fairly constant~\cite{meyer2020}. It was shown for fcc-Rh/Co/Ir(111) that the conical spin spirals, confined close to the FM state, produce an energy dispersion similar to that of the flat spin spirals, with magnetic moments of the Rh and Co layers remain constant throughout the two high-symmetry directions. Consequently, the exchange interaction constants evaluated by mapping the energies of the flat and conical spin spirals onto the Heisenberg Hamiltonian are close to each other, indicating that the exchange constants are independent of the variations in the magnetic moments for the film~\cite{meyer2020}. 

For the three films considered here, the magnetic moments of the $4d$ overlayers, i.e., Rh, Pd, and Ru, and the Co layers remain fairly constant in the vicinity of the FM state (Figs.~\ref{fig:f2}(a-c)). Similar to the Rh/Co/Ir(111) film, the exchange constants for the three films can be regarded as independent of the variation in magnetic moments.

Subsequently, we fit the energy dispersion of spin spirals calculated without and with SOC to the Heisenberg
(Eq.~(1)) and DMI (Eq.~(2)) Hamiltonians, respectively, and extract the interaction constants. The HOI constants are evaluated from the energy differences between the multi-$Q$ and single-$Q$ states (Eqs.~(6-11)). All these interaction constants for the three films are listed in Table~\ref{tab:table1}. The effective exchange ($J_{\mathrm{eff}}$) and DMI ($D_{\mathrm{eff}}$) constants are obtained by fitting the energy dispersion in the vicinity of the $\overline{\Gamma}$ point (insets of Figs.~\ref{fig:f1}(b-d) and Figs.~\ref{fig:f1}(e-g)). These two effective interaction constants, along with the MAE, for all three films, are taken from Ref.~\cite{paul2024}.

First, we focus on the insets of Figs.~\ref{fig:f1}(b-d), where the energy dispersion of the spin spirals excluding and including SOC around the $\overline{\Gamma}$ point ($\mid$$\textbf{q}$$\mid \leq 0.1\times\frac{2\pi}{a}$) is displayed for three films. The behavior of the energy dispersion can be interpreted by an effective spin model containing the effective exchange interaction ($J_{\mathrm{eff}}$), effective DMI ($D_{\mathrm{eff}}$), and MAE. As described above, the effective exchange interaction ($J_{\mathrm{eff}}$) captures the parabolic nature of the dispersion curve, excluding SOC, close to the $\overline{\Gamma}$ point. The out-of-plane (easy axis) anisotropy favors the FM alignment, in which all spins point along the out-of-plane direction ($z$-axis). However, within a spin spiral, the orientation of the spin rotates gradually from the positive to the negative $z$-axis. As a result, the effect of MAE on the dispersion of the spin spiral is a shift in energy with respect to the FM state by $K_{\mathrm{MAE}}/2$.

The positive value of the effective exchange constant ($J_{\mathrm{eff}}$) indicates that the ground state of all three films is ferromagnetic. The magnitude of the constant is a measure of the curvature around the $\overline{\Gamma}$ point. The effective constant of fcc-Ru/Co/Re(0001) is nearly two times smaller than those of hcp-Pd/Co and fcc-Rh/Co on Re(0001). This indicates that fcc-Ru/Co/Re(0001) exhibits the smallest curvature, while hcp-Pd/Co/Re(0001) shows the highest curvature, and that value of fcc-Rh/Co/Re(0001) lies in between. Note that the effective exchange constant of fcc-Rh/Co ($J_{\mathrm{eff}} \approx$ 5.4 meV) and hcp-Rh/Co ($J_{\mathrm{eff}} \approx$ 6.0 meV) on Ir(111) is even smaller than that of fcc-Ru/Co/Re(0001)~\cite{meyer19}.

The positive value of the effective DMI constant ($D_{\mathrm{eff}}$) in all three films indicates that the right-rotating, i.e., clockwise, cycloidal spin spirals are favored. The value of the DMI constant for fcc-Rh/Co/Re(0001) is largest among the three considered film systems (Table~\ref{tab:table1}). However, the MAE counterbalances the energy gain by DMI and stabilizes the FM configuration as the ground state. 

The effective exchange constant ($J_{\mathrm{eff}}$) of hcp-Pd/Co/Re(0001) is nearly 20\% larger than that of fcc-Rh/Co/Re(0001), which makes the dispersion curve rise more quickly around the $\overline{\Gamma}$ point. At the same time, the effective DMI constant ($D_{\mathrm{eff}}$) is nearly 40\% lower than that of fcc-Rh/Co/Re(0001). Since the MAE is still sufficiently large, no spin spiral ground states are found in hcp-Pd/Co/Re(0001). On the other hand, the effective exchange constant of fcc-Ru/Co/Re(0001) is lowest among the three films, leading to a very flat dispersion curve around the $\overline{\Gamma}$ point (inset of Fig.~\ref{fig:f1}(d)). However, due to the smallest effective DMI constant, in fcc-Ru/Co/Re(0001), and considerable MAE, no spin spiral states become energetically favorable than the FM state. 

The MAE of hcp-Pd/Co and fcc-Ru/Co on Re(0001) is approximately 4 to 5 times smaller than fcc-Rh/Co/Re(0001), causing a minimal change to the energy dispersion, and the FM state remains the lowest energy configuration for those two films. The effective DMI constant of hcp-Rh/Co/Ir(111) ($D_{\mathrm{eff}}=$ 1.64 meV) is nearly 1.4 times larger than that of fcc-Rh/Co/Re(0001), whereas the value for fcc-Rh/Co/Ir(111) ($D_{\mathrm{eff}}=$ 0.83 meV) is close to the value of hcp-Pd/Co/Re(0001). The MAE for both the Rh/Co bilayers on Ir(111) is approximately 2.7 times larger than that of fcc-Rh/Co/Re(0001)~\cite{meyer19}.  

It was shown in Ref.~\cite{paul2024} for selected films, including fcc-Rh/Co/Re(0001), that the effective exchange interaction explains the energy dispersion curve close to the $\overline{\Gamma}$ point very well. However, it fails to describe the dispersion curve at moderate and large $q$ values, where an extended model is required. The model includes the beyond nearest-neighbor exchange interactions which are responsible for the deviation of the dispersion curve from its parabolic nature. For this reason, we calculate the beyond nearest-neighbor exchange interaction constants for hcp-Pd/Co and fcc-Ru/Co on Re(0001). 

The nearest-neighbor exchange constant ($J_1$) is a rough measure of the energy difference between the FM and AFM states. Naturally, this quantity is poorly estimated within the effective model, since it only explains the energy dispersion close to the $\overline{\Gamma}$ point. A comparison of this exchange constant with the effective one reveals that $J_{\mathrm{eff}}$ is 33\% smaller for fcc-Rh/Co/Re(0001), while its value is 44\% and 28\% higher for hcp-Pd/Co and fcc-Ru/Co on Re(0001), respectively. In the case of fcc-Rh/Co and hcp-Rh/Co on Ir(111), $J_{\mathrm{eff}}$ is nearly 370\% smaller than $J_1$~\cite{meyer19}. The negative signs of the beyond nearest-neighbor exchange constants for the three films indicate that they favor antiparallel alignment of spins, and thus induce exchange frustration into the system (Table~\ref{tab:table1}).         

The effective DMI constant only describes the linear portion of the energy dispersion curve of the cycloidal spin spiral close to the $\overline{\Gamma}$ point (Figs.~\ref{fig:f1}(e-g)). But the dispersion curve deviates from a linear dependence at moderate and large $q$ values, which indicates the presence of the beyond nearest-neighbor terms and therefore, an extended spin model is required. The DMI constant up to the fifth nearest-neighbor is listed in Table~\ref{tab:table1}. The negative signs of the DMI constants indicate the presence of frustration, which further flattens the energy dispersion curve around the $\overline{\Gamma}$ point, as seen in the insets of Figs.~\ref{fig:f1}(b-d).

The effective DMI constant, $D_{\mathrm{eff}}$, is smaller than the nearest-neighbor constant ($D_1$) for all three films considered here. The effective constant ($D_{\mathrm{eff}}$) is nearly 25\% and 70\% smaller than $D_1$ for fcc-Rh/Co and hcp-Pd/Co on Re(0001), respectively. For fcc-Ru/Co/Re(0001), the value of $D_{\mathrm{eff}}$ is extremely small, i.e., 0.07 meV, compared to the other two films, and the value of $D_1$ is 0.55 meV. For comparison, we also note the effective ($D_{\mathrm{eff}}$) and nearest-neighbor DMI ($D_1$) constants of hcp-Rh/Co and fcc-Rh/Co on Ir(111). The value of $D_{\mathrm{eff}}$ for fcc-Rh/Co/Ir(111) is 0.83 meV. However, the value of $D_1$ becomes smaller (0.29 meV), and it shows an opposite (negative) sign. In case of hcp-Rh/Co/Ir(111), $D_{\mathrm{eff}}$ is 1.63 meV and $D_1$ is only 0.13 meV.   

The three HOI constants, namely, biquadratic ($B_1$), three-site four spin ($Y_1$) and four-site four spin ($K_1$), for the three films are evaluated from the energy difference between the multi-$Q$ and single-$Q$ states using Eqs.~(6-8). The energy difference and the three HOI constants for the three films are listed in Table~\ref{tab:table1}.

The three-site four spin ($Y_1$) and four-site four spin ($K_1$) interaction constants for fcc-Rh/Co and hcp-Pd/Co on Re(0001) are negative, while the biquadratic constant ($B_1$) is positive for these two films. The values of the four-site four spin ($K_1$) and the biquadratic ($B_1$) constants differ by an order of magnitude between the two films, whereas the values of the three-site four spin ($Y_1$) constants remain comparable. Note that fcc-Ru/Co/Re(0001) has vanishing three-site and four-site four spin interaction constants. However, the biquadratic constant has a finite positive value, comparable to that of fcc-Rh/Co/Re(0001). 

We notice that the HOI constants of the Rh/Co and Pd/Co bilayers on the Ir(111) possess the same signs as those of the fcc-Rh/Co and hcp-Pd/Co bilayers on Re(0001)~\cite{gutzeit2021}. The value of the four-site four spin constant ($K_1$) for fcc-Pd/Co and hcp-Rh/Co on Ir(111), is 1.41 and 1.01 meV, respectively, whereas it is nearly by a factor of two smaller for fcc-Rh/Co/Ir(111), i.e., 0.58 meV. The value of the three-site four spin interaction constant ($Y_1$) of these three Ir-based films varies between 0.96 and 1.55 meV. The value of the biquadratic interaction constant ($B_1$) of fcc-Pd/Co and fcc-Rh/Co on Ir(111) is 1.60 and 1.79 meV, respectively, comparable to that of hcp-Pd/Co/Re(0001), while the value is one order of magnitude smaller for hcp-Rh/Co/Ir(111), i.e., 0.34 meV. It is important to note that, in contrast to the Co-based films, the three-site four spin and four-site four spin constants are mostly positive in Fe-based films~\cite{gutzeit2021}.

\begin{figure*}[!htbp]
	\includegraphics[scale=1.0]{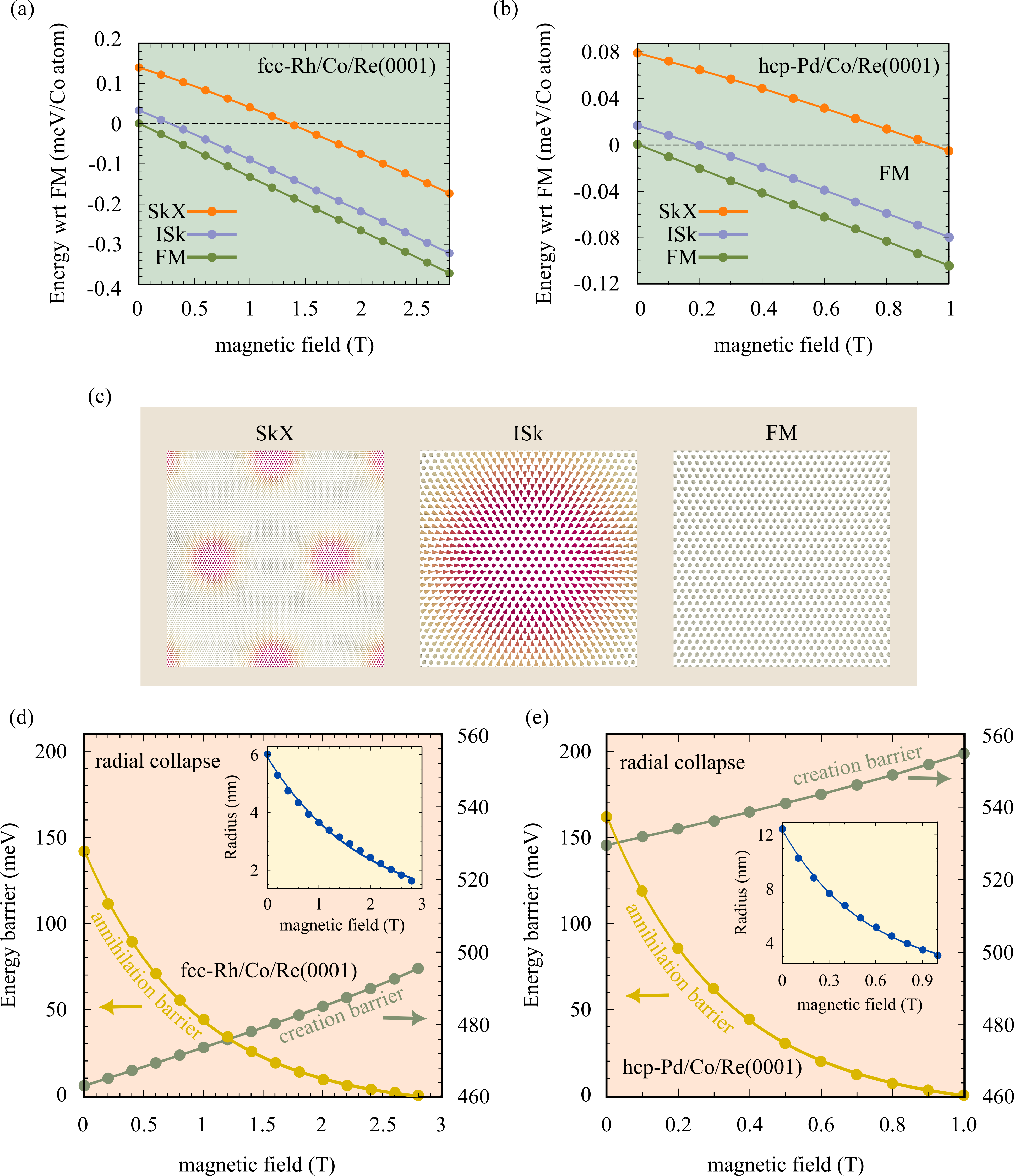}
	\centering
	\caption{\justifying Zero-temperature magnetic phase diagram of (a) fcc-Rh/Co/Re(0001), and (b) hcp-Pd/Co/Re(0001). Energies of the FM state (green), skyrmion lattice (SkX, orange), and isolated skyrmions (ISk, blue) are shown as a function of the external magnetic field. (c) Spin structure of the skyrmion lattice (SkX), an isolated skyrmion, and the FM state for fcc-Rh/Co/Re(0001) at zero magnetic field. Only a small portion of the $150 \times 150$ simulation box is shown. Annihilation (yellow) and creation (green) barriers of isolated skyrmion for (d) fcc-Rh/Co/Re(0001) and (e) hcp-Pd/Co/Re(0001) as a function of applied magnetic field obtained via the GNEB method. The solid circles are obtained from the atomistic spin model (Eq.~(13)), the annihilation barriers ($E^{a}_b$) are fitted with $E^{a}_b(B)= a \left(B + b\right)^{-1} + c$ and the creation barriers ($E^{c}_b$) with $E^{c}_b(B)= mB + c$, where $a$, $b$, $c$, and $m$ are fitting parameters and $B$ denotes the magnetic field. The inset shows variation in the radius of isolated skyrmions for (d) fcc-Rh/Co/Re(0001) and (e) hcp-Pd/Co/Re(0001) with the magnetic field. The solid circles are obtained from the atomistic spin model (Eq.~(13)) and solid lines are fit to $R(B)= a\left(B + b \right)^{-1} + c$, where $R$ denotes the radius, $B$ denotes the magnetic field and $a$, $b$, and $c$ are the fitting parameters. The values of the fitting parameters are listed in Table~\ref{tab:table2} of Appendix~\ref{sec:appnA}.}
	\label{fig:f3}
\end{figure*}

\begin{figure*}[!htbp]
	\includegraphics[scale=1.0]{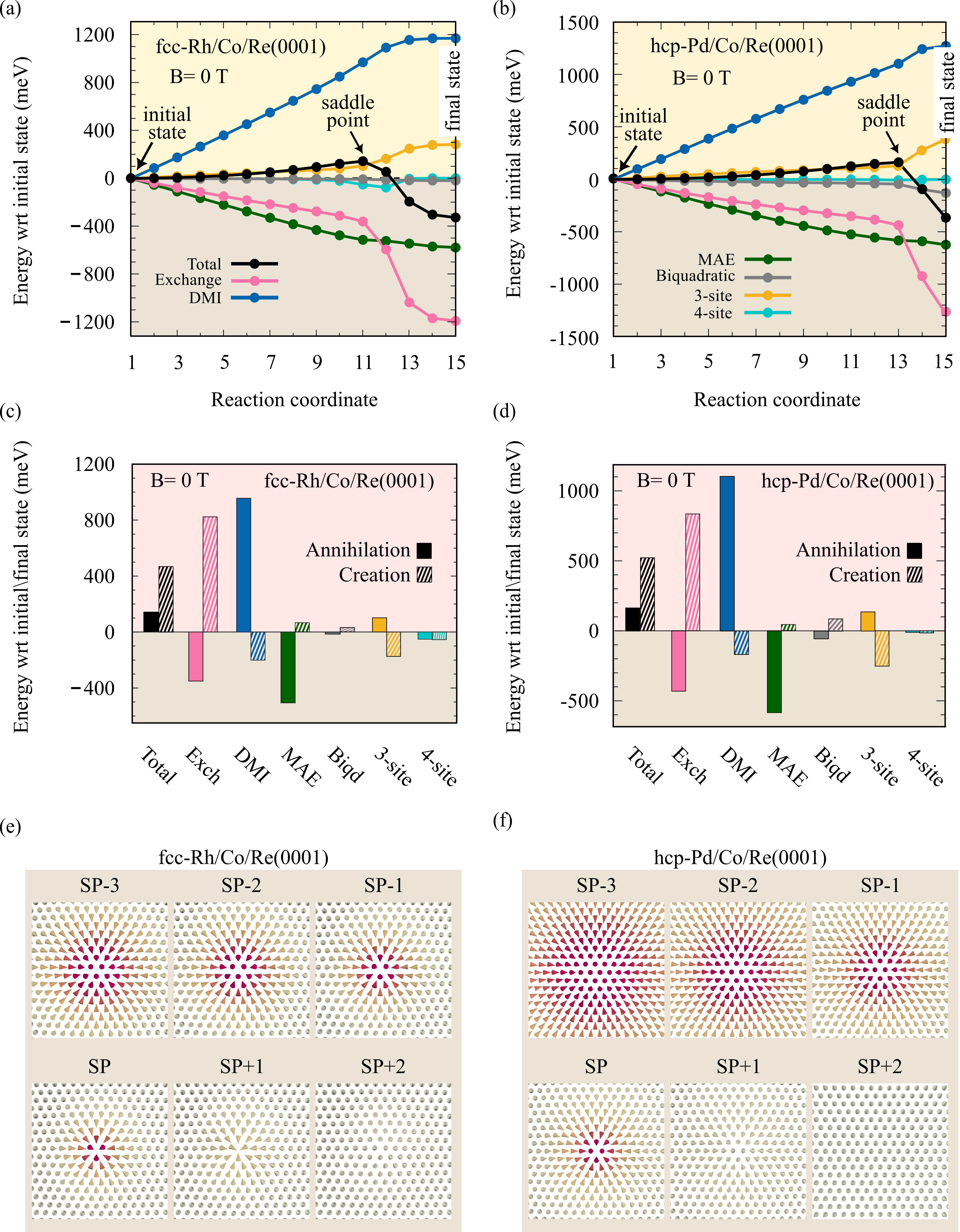}
	\centering
	\caption{\justifying Total and individual energy contributions with respect to the initial (skyrmion) state along the minimum energy path (MEP) for the radial collapse of isolated skyrmions into the FM (final) state at zero magnetic field for (a) fcc-Rh/Co/Re(0001) and (b) hcp-Pd/Co/Re(0001). Energy of the saddle point with respect to the initial (skyrmion) and final (FM) states defines the annihilation and creation barriers, respectively. Total and individual energy decomposition of the annihilation and creation barriers for (c) fcc-Rh/Co/Re(0001) and (d) hcp-Pd/Co/Re(0001). Three-site four spin and four-site four spin interactions are abbreviated as 3-site and 4-site, respectively. Spin structures before (SP-3, SP-2, and SP-1), after (SP+1 and SP+2), and at the saddle point (SP) for (e) fcc-Rh/Co/Re(0001) and (f) hcp-Pd/Co/Re(0001) along the MEP shown in (a) and (b), respectively, indicating radial collapse. Note that only a small portion around the skyrmions in the $150 \times 150$ simulation box is shown.} 
	\label{fig:f4}
\end{figure*}

It is clear from the above analysis that the extended spin model (Eq.~(13)) -- which contains the beyond nearest-neighbor exchange interactions, the beyond nearest-neighbor DMI, MAE, and HOI, and the effective model~\cite{paul2024} -- which contains the effective exchange interaction, effective DMI, and MAE -- predict the ground state of the three films correctly. However, the ultrathin films investigated here possess significant amount of frustration, arising from competing beyond nearest-neighbor terms, which is not captured by the effective spin model. Furthermore, the exchange frustration has been shown to enhance the energy barriers of isolated skyrmions in transition-metal based ultrathin films~\cite{malottki2017a}. Consequently, for an accurate assessment of the skyrmion stability in our film systems, the beyond nearest-neighbor terms need to be taken into account. Therefore, the extended spin model is best suited for the study of stability of the metastable skyrmions in our film systems. In the following sections, we explore the possibility of stabilizing isolated skyrmions in these film systems using atomistic spin simulations based on the extended spin model with parameters obtained from DFT.

\section{Atomistic spin simulations}
\subsubsection{Zero-temperature phase diagram}

We study the relaxation of spins by treating them as classical objects using the Landau-Lifshitz equation (Eq.~(12)) involving an atomistic spin Hamiltonian (Eq.~(13)) completely parameterized from DFT. In particular, we first calculate the zero-temperature phase diagram to verify our prediction that isolated skyrmions can be stabilized without an external magnetic field. Subsequently, we focus on calculating skyrmion properties, such as their radius, collapse mechanism, and energy barriers, which allow us to estimate their stability.

We take various starting spin configurations, such as the FM state, skyrmion lattice (SkX), and isolated skyrmions (ISk), and relax them using the Landau-Lifshitz equation in the presence of external magnetic fields applied perpendicular to the film. Note that the skyrmions are created on the FM background 
using the theoretical profile described in Ref.~\cite{bogdanov1994b}. The energies of the three relaxed spin structures as a function of the magnetic field are shown for fcc-Rh/Co and hcp-Pd/Co on Re(0001) in Figs.~\ref{fig:f3}(a,b), respectively.

In the zero-temperature phase diagram, we observe that the FM configuration is the ground state for the whole range of magnetic field for both films (Figs.~\ref{fig:f3}(a,b)). As predicted in the previous section, isolated skyrmions (ISk) appear spontaneously as metastable states on the FM ground state at zero magnetic field for both films. The energy of isolated skyrmions remains slightly above the FM ground state throughout the range of the magnetic field, whereas the relaxed skyrmion lattice (SkX) remains noticeably higher in energy than the isolated skyrmions. The relaxed spin structures of the skyrmion lattice (SkX), isolated skyrmions (ISk), and the FM state at zero magnetic field for fcc-Rh/Co/Re(0001) are shown in Fig.~\ref{fig:f3}(c). The spin structures of the isolated skyrmions and skyrmion lattice are similar to those obtained in other TM ultrathin films~\cite{malottki2017a,som18,meyer19,Muckel2021,paulhoi,paul2022}.

It is important to note that the FM state remains the ground state at zero and finite magnetic fields for fcc-Ru/Co/Re(0001). However, isolated skyrmions could not be stabilized at zero or finite magnetic fields. This occurs due to the low value of the DMI in this film. Therefore, we do not show its phase diagram.

\subsubsection{Radius and stability of isolated skyrmions}

We calculate the radius of isolated skyrmions in fcc-Rh/Co and hcp-Pd/Co on Re(0001) as a function of the magnetic field based on the definition by Bogdanov~\cite{bogdanov1994b} (inset of Figs.~\ref{fig:f3}(d,e), respectively). The radius of isolated skyrmions at zero field is 12.4 nm for hcp-Pd/Co/Re(0001) and 6~nm for fcc-Rh/Co/Re(0001), and it reduces to a value of 3.6 nm at 1~T and 1.6 nm at 2.8~T, respectively. Note that isolated skyrmions with an even smaller radius of about 2.5 nm were observed experimentally via SP-STM measurements in Rh/Co/Ir(111) at zero magnetic field~\cite{meyer19}.

Next, we study the collapse mechanism of isolated skyrmions (initial state) into the FM ground state (final state) by calculating the minimum energy path (MEP) using the geodesic nudged elastic band (GNEB) method~\cite{bessarab2015}. The energy difference between the saddle point -- the maximum energy point on the MEP -- and the initial state, is defined as the annihilation barrier, whereas the energy difference between the saddle point and the final state is defined as the creation barrier. The creation and annihilation barriers as a function of the magnetic field for fcc-Rh/Co and hcp-Pd/Co on Re(0001) are shown in Figs.~\ref{fig:f3}(d,e), respectively.

The annihilation barriers of isolated skyrmions for the two films follow the same trend as the radius with the magnetic field. This standard behavior of the annihilation barriers is observed in various TM ultrathin films~\cite{malottki2017a,som18,paulhoi,paul2022}. The annihilation barriers of fcc-Rh/Co/Re(0001) are about 140 meV, while those of hcp-Pd/Co/Re(0001) are about 160 meV at zero magnetic field. These annihilation barriers decrease with increasing magnetic field and almost vanishes at 2.8~T and 1~T for fcc-Rh/Co/Re(0001) and hcp-Pd/Co/Re(0001), respectively. They decrease with the magnetic field, since all the interactions scale with the size of skyrmions.

Experimentally, isolated skyrmions with diameters of a few nanometers were observed via SP-STM measurements in Pd/Fe/Ir(111) under magnetic field~\cite{Romming2013,Romming2015,Muckel2021}. The annihilation barriers of isolated skyrmions in this ultrathin film are on the order of 180 meV~\cite{meyer19,malottki2017a,paulhoi}. Therefore, the calculated energy barriers of skyrmions in fcc-Rh/Co and hcp-Pd/Co on Re(0001) are large enough that they could be identified in low temperature SP-STM experiments. 

The creation barriers, on the other hand, increase linearly by a small amount with magnetic field for the two considered films. Their values are about 460 meV for fcc-Rh/Co/Re(0001) and almost 530 meV for hcp-Pd/Co/Re(0001) at zero magnetic field and increasing to almost 500 meV at 2.8~T and nearly 560 meV at 1~T, respectively. The creation barriers arise as a competition between the annihilation barriers and the energy difference between the initial and final states. The annihilation barriers decrease, and the energy difference between the initial and final states increases with the magnetic field. The creation barrier is the result of the two opposite effects, which follow the same trends as reported in Ref.~\cite{paul2022}.

Next, we analyze the minimum energy path for the collapse of an isolated skyrmion into the FM ground state at zero magnetic field for the two films. For this purpose, we decompose the total energy along the minimum energy path into various contributing energy terms, which are shown in Figs.~\ref{fig:f4}(a,b) for fcc-Rh/Co and hcp-Pd/Co on Re(0001), respectively.

We observe that isolated skyrmions annihilate via the well-known radial collapse mechanism~\cite{Muckel2021} for both films (Figs.~\ref{fig:f4}(e,f)). In the process, the isolated skyrmion shrinks radially along the minimum energy path up to the saddle point. Close to the saddle point, the central spins of skyrmions begin to rotate towards the in-plane direction. The saddle point is the critical configuration where the central spins achieve maximum rotation towards the in-plane direction without breaking the topological protection. Further rotation of these spins beyond this critical point destroys the topological protection, after which the spins reorient to the FM order, completing the annihilation process.

The individual interaction terms along the minimum energy path behave in a similar manner for both films (Figs.~\ref{fig:f4}(a,b)). The exchange, DMI, MAE, and three-site four spin interaction terms vary almost linearly from the initial state to the saddle point along the minimum energy path as the size of the isolated skyrmions reduces. 

The exchange interaction and MAE favor the FM (final) state over the isolated skyrmion (initial state), therefore, decrease along the MEP. On the other hand, the DMI and three-site four spin interaction ($Y_1$) favor skyrmions over the FM state, and increase along the MEP. Due to the large spin orientation around the saddle point, the exchange and three-site four spin interaction terms change significantly. The DMI and MAE terms also feel this rapid orientation of spins, but due to their smaller interaction constants, the changes are comparatively small. In the end, the spins are reoriented by small amounts to align themselves to the FM configuration, which leads to a flat region.

The four-site four spin interaction ($K_1$) term for fcc-Rh/Co/Re(0001) behaves in a similar manner along the MEP as reported in Ref.~\cite{paulhoi} for skyrmions in Fe-based ultrathin films. In particular, it only attains a significant value close to the saddle point, where the angles among the central spins of skyrmions are large. However, since the sign of the four-site four spin interaction is negative, it lowers the annihilation barrier, i.e., reduces the stability of skyrmions in this film. The contribution from the four-site four spin interaction nearly vanishes for hcp-Pd/Co/Re(0001), since the constant is about 4 times smaller than that of fcc-Rh/Co/Re(0001). The contribution of the biquadratic ($B_1$) term is negligible for both films.

To understand the creation and annihilation energy barriers in more detail, we decompose the saddle point at zero field (Figs.~\ref{fig:f4}(a,b)) into constituting energy terms and show in Figs.~\ref{fig:f4}(c,d) for the two films. The DMI makes a large positive contribution, while the three-site four spin ($Y_1$) term has a relatively small positive contribution to the annihilation barrier, thereby increasing the stability of isolated skyrmions. In contrast, the exchange and MAE terms provide a negative contribution to the energy barrier, since they favor the FM (final) state and therefore, reduce the stability of skyrmions. The biquadratic ($B_1$) and the four-site four spin ($K_1$) interactions have negligible contributions to the energy barrier.

The contributions of the interaction terms change sign for the creation barriers. Now, the DMI and three-site four spin ($Y_1$) terms have negative contributions, and therefore decrease the creation barrier, whereas the exchange interaction term makes a positive contribution, and thus increases the energy barrier. The contributions of the other interactions are negligible.

\section{\label{sec:conc} Conclusion}

Using an atomistic spin model, we have explored the possibility of stabilizing isolated skyrmions at zero and finite magnetic fields in fcc-Rh/Co, hcp-Pd/Co, and fcc-Ru/Co atomic bilayers on the Re(0001) surface. The spin model contains the Heisenberg pairwise exchange interaction, Dzyaloshinskii-Moriya interaction, magnetocrystalline anisotropy energy, as well as the beyond Heisenberg multi-spin higher-order interactions. All the interaction parameters of the spin model are obtained from DFT.

The calculated magnetic phase diagram shows that isolated skyrmions emerge spontaneously on the ferromagnetic ground state in fcc-Rh/Co and hcp-Pd/Co on Re(0001) even in the absence of a magnetic field, while skyrmions cannot be stabilized in fcc-Ru/Co/Re(0001) due to the small Dzyaloshinskii-Moriya interaction. The radius of isolated skyrmions in fcc-Rh/Co/Re(0001) is nearly 6 nm at zero magnetic field, whereas it is about 12 nm in hcp-Pd/Co/Re(0001).

The annihilation barriers, which protect isolated skyrmions from collapsing into the ferromagnetic background, are on the order of 150 meV and thereby sufficiently high, such that skyrmions may be realized experimentally in SP-STM measurements at low temperatures. Our analysis of the energy barriers reveals that the Dzyaloshinskii-Moriya interaction and the three-site four spin interaction stabilize the skyrmions, while the pairwise Heisenberg exchange interaction, biquadratic interaction, and magnetocrystalline anisotropy energy terms act in the opposite direction. The four-site four spin interaction is small in these systems and provides only a negligible contribution. 

In summary, from our study 4$d$/Co atomic bilayers on the Re(0001) surface appear a promising platform for realizing isolated skyrmions with diameters on the order of 20 nm and below at zero magnetic field, which may trigger future experiments. Furthermore, these films have the potential to become model magnet-superconductor hybrid systems, since the Re substrate becomes superconducting below 1.7~K.

\section{\label{sec:ackn} Acknowledgments}

The authors acknowledge the computing time on ``Lise" at the NHR center NHR@ZIB and on ``Emmy" at the NHR center NHR@G\"{o}ttingen. These centers are jointly supported by the Federal Ministry of Education and Research and the state governments participating in the NHR~\cite{nhr}. S.P. graciously acknowledges IISER Thiruvananthapuram for funding and computing time on the Padmanabha cluster. S.P. also acknowledges funding from the Anusandhan National Research Foundation (ANRF/ECRG/2024/001865/PMS).

\appendix

\begin{table*}[!thbp]
	\centering
	\caption{\justifying Parameters $a$, $b$, and $c$ of the radius (R) versus magnetic field (B) fitting function $R(B)= a \left( B + b \right)^{-1} + c$, parameters $m$ and $c$ of the creation barriers ($E^{c}_b$) versus magnetic field (B) fitting function $E^{c}_b(B)= mB + c$ and parameters $a$, $b$, and $c$ of the annihilation barriers ($E^{a}_b$) versus magnetic field (B) fitting function $E^{a}_b(B)= a \left( B + b \right)^{-1} + c$ used in Fig.~\ref{fig:f3}.} 
	\label{tab:table2}
	\begin{ruledtabular}
		\begin{tabular}{c *{3}{c} *{2}{c} *{3}{c}} % Using 'c' for general compatibility in ruledtabular
			\begin{tabular}{@{}c@{}}
				\\
				Systems
			\end{tabular} & \multicolumn{3}{c}{Radius (nm)} & \multicolumn{2}{c}{\hspace*{-0.8em}{Creation barrier (meV)}} & \multicolumn{3}{c}{\hspace*{-1em}{Annihilation barrier (meV)}} \\
			\cmidrule{2-4} \cmidrule{5-6} \cmidrule{7-9}
			\noalign{\vskip+0.6ex}
			& $a$ & $b$ & $c$ & {\hspace*{+1em}}$m$ & {\hspace*{+2em}}$c$ & {\hspace*{+2em}}$a$ & {\hspace*{+1em}}$b$ & {\hspace*{-2em}}$c$ \\
			& $\textrm{(nm\,.T)}$ & $\textrm{(T)}$ & $\textrm{(nm)}$ & {\hspace*{+1em}}$\textrm{(meV/T)}$ & {\hspace*{+2em}}$\textrm{(meV)}$ & {\hspace*{+2em}}$\textrm{(meV\,.T)}$ & {\hspace*{+1em}}$\textrm{(T)}$ & {\hspace*{-2em}}$\textrm{(meV)}$ \\
			\colrule % Separating header from data
			fcc-Rh/Co/Re(0001) &67.94 &2.41 &$-$6.82 &11.20 &464.17 &186.78 &0.97 &$-$51.55 \\
			hcp-Pd/Co/Re(0001) &39.09 &0.69 &$-$11.82 &24.58 &529.21 &81.15 &0.36 &$-$60.29 \\
		\end{tabular}
	\end{ruledtabular}
\end{table*}

\section{\label{sec:appnA} Fitting parameters}

We fitted the variation of radius (R) with external magnetic field ($B$) in the insets of Figs.~\ref{fig:f3}(d) and \ref{fig:f3}(e) for fcc-Rh/Co/Re(0001) and hcp-Pd/Co/Re(0001), respectively, using $R(B)= a \left( B + b \right)^{-1} + c$, where $a$, $b$, and $c$ are the fitting parameters. Similarly, we also fitted the variation of the creation ($E^{c}_b$) and annihilation ($E^{a}_b$) barriers with magnetic field ($B$) using $E^{c}_b(B)= mB + c$ and $E^{a}_b(B)= a \left( B + b \right)^{-1} + c$, respectively, where $a$, $b$, $c$, and $m$ are fitting parameters. These fits were performed for both fcc-Rh/Co/Re(0001) and hcp-Pd/Co/Re(0001) films and are displayed in Figs.~\ref{fig:f3}(d) and \ref{fig:f3}(e), respectively. The values of these fitting parameters are listed in Table~\ref{tab:table2}.

\end{document}